# Quantization-aware Photonic Homodyne computing for Accelerated Artificial Intelligence and Scientific Simulation

*Lian Zhou[1,2,3]\*, Kaiwen Xue[1]\*, Amirhossein Fallah[3]\*, Lijin Liu[1]\*, Chun-Ho Lee[1,2], Kiwon Kwon[1], Clayton Cheung[1], Yuan Li[2], Yue Yu[1], Yun-Jhu Lee[2], Songlin Zhao[1], Ryan Hamerly[2,4], Edo Waks[5], Dirk Englund[4], Constantine Sideris[3,6]†, Mengjie Yu[1,2,3]‡, and Zaijun Chen[1,2,3]#*

[1]*Department of Electrical Engineering and Computer Science, University of California, Berkeley, CA 94720, USA*
[2]*Opticore Inc., Berkeley, CA, 94704, USA*
[3]*Department of Electrical and Computer Engineering, University of Southern California, Los Angeles, CA 90089, USA*
[4]*Research Lab of Electronics, MIT, Cambridge, MA 02139*
[5]*Department of Electrical and Computer Engineering, University of Maryland, College Park, MD 20742, USA*
[6]*Department of Electrical Engineering, Stanford University, Stanford, CA 94305*
\**These authors contributed equally*
E-mail: zaijun@berkeley.edu, mengjie.yu@berkeley.edu, csideris@usc.edu

**Abstract**

Modern problems in high-performance computing, ranging from training and inferencing deep learning models in computer vision and language models to simulating complex physical systems with nonlinearly-coupled equations, require exponential growth of computational resources. Photonic analog systems are emerging with solutions of intrinsic parallelism [1–4], high bandwidth [5–9], and low propagation loss [10–12]. However, their application has been hindered by the low analog accuracy due to the electro-optic distortion, material nonlinearities, and signal-to-noise ratios. Here we overcome this barrier with a quantization-aware digital-photonic mixed-precision framework across chiplets for accelerated AI processing and physical simulation. Using Lithium Niobate photonics with channel equalization techniques, we demonstrate linear multiplication (9-bit amplitude-phase decoupling) in homodyne optical logics with 6-bit precision at the clock rate of 128 giga-symbol-per-second (128 GS/s), enabling AI processing with 6 ns latency. Codesign hardware-algorithms, including iterative solvers, sparse-dense quantization, and bit-sliced matrix multiplication, explore photonic amplitude and phase coherence for complex-valued, physics-inspired computation. In electromagnetic problems, our approach yields 12-bit solutions for partial differential equations (PDEs) in scattering problems that would conventionally require up to 32-bit and often even 64-bit precision. These results preserve digital-level fidelity while leveraging the high-speed low-energy photonic hardware, establishing a pathway toward general-purpose optical acceleration for generative artificial intelligence [16], real-time robotics [17], and accurate simulation for climate challenges [18] and biological discoveries [19].

## Introduction

Computational science, by simulating physical reality on computers, allows researchers to design, predict, and optimize complex systems across disciplines, which dramatically accelerates discoveries from nanoscale materials modeling [20], molecular dynamics [19], and semiconductor device design [21], to global-scale climate prediction and earth-system modeling [18]. These systems, such as electromagnetic simulation for radar cross-sections (RCSs), are rarely tractable by analytical techniques, and direct numerical computation is hampered by the exponential growth in resource requirements with increasing model size and complexity. In parallel, the rapid expansion of artificial intelligence (AI) is revolutionizing science and technology. However, training and deploying large deep learning models has placed a pressing strain on high-performance computing (HPC) infrastructure, demanding both vast energy costs and extreme throughput. Conventional CMOS-based architectures are increasingly inadequate for this challenge, specifically for data-intensive algorithms. As Moore's law slows down and Dennard scaling collapses, improvements in transistor density no longer translate into proportional gains in performance or energy efficiency. The resulting bottlenecks in clock speed, power consumption, and data movement impose unsustainable costs for data-intensive computing tasks. Moreover, complex numbers are the native language of waves, and thus are essential for electromagnetics and quantum mechanics, where solvers are operated on complex fields and simulated with iterative linear-algebra kernels, such as FFT-based Helmholtz/Poisson solvers [22], scattering and inversion [23], non-Hermitian systems [24], all of which require complex multiply–accumulate at scale. In machine learning, complex activations provide a fundamental building block that can solve nonlinear problems, such as Exclusive-OR (XOR) tasks [25] using a single-layer perception, which was otherwise impossible with all real-valued operators. Recent work has also shown that complex-valued data in neural networks provides richer representations, faster convergence, and stronger generalization [26,27]. However, digital electronic platforms are inherently real-valued and cannot operate on complex numbers directly. Complex numbers are multiplied using four real-number multiplications with two additions, leading to significant overhead.

Leveraging the wave property of light, photonic computing uses data movement with dielectric waveguides without capacitive resistance, which is crucial for high clock rates and low propagation loss [2,11,28,29], and its amplitude-phase coherence allows for direct complex-valued operations [2][30]. However, as an analog system, the accuracy in optical systems remains constrained, especially at high clock rate operations, due to device and material nonlinearity such as carrier dispersion in silicon modulators, parasitic noises and distortions in electro-optic (EO) conversion, and analog signal-to-noise ratios. So far, high-speed systems (with clock rates over 10 GS/s) achieved only about 3-5 bits of accuracy (Table 2) [3,5,6,31,32]. As a result, photonic computing has been mostly benchmarked as accelerators for fault-tolerant tasks [6,33,34], limiting its flexibility for complicated AI tasks and scientific problems. Moreover, existing photonic computing approaches based on space or wavelength mapping required $O(N^2)$ weighting modulators [1,4,5,34–36] to encode a matrix, which doesn't scale well due to the limited chip area, large photonic device footprint, and fabrication variations. Recently, time-multiplexed computing systems, leveraging weight mapping in time steps, have emerged to reduce the modulator counts from $O(N^2)$ to $O(N)$ [6–8,10,32,37,38], however, its realization with cascading intensity modulation [3,6–8,10,32] requires wavelength multiplexing to amortize the energy cost but that leads to $O(N^2)$ de-multiplexers, hampering its scalability [3,7,10,39,40]. Alternatively, homodyne logics [37] with instantaneous photoelectric multiplication between two laser fields with compact coherent detectors enable spatial parallelism using free-space fanout and on-chip beam routing [37,41] toward high channel counts, but due to the amplitude-phase coupling with semiconductor lasers [38] or thermal-optic silicon photonic MZIs [42], as well as carrier-based silicon modulators, a linear homodyne compute unit has not been demonstrated yet.

Here we propose a quantization-aware approach to enable high precision tasks in optical computing based on advancements on (1) architectural-level mixed-precision framework (Fig. 1a and 1b) that maximizes the throughput using a low-precision optical processor (7-8 bits) for data-heavy operations, while maintaining the model fidelity (12 bits), noting that these approaches are emerging and applicable in state-of-the-art AI algorithms (e.g., SqueezeNet [43], DeepSeek [44], and Hardware-aware Automated Quantization [45]); (2) device and material level with the first linear homodyne photoelectric multiplication logic to scale up optical computing, based on the amplitude- or phase-modulation in thin-film Lithium niobate (TFLN) electro-optics that allows simultaneously high-speed (128 GS/s), low-voltage and low-propagating-loss operations; (3) channel equalization techniques to eliminate high-speed EO response distortions for accuracy improvement and (4) physics-inspired photonic complex-valued computing for solving PDEs in physical simulations.

## Results

### 2.1. System Architecture

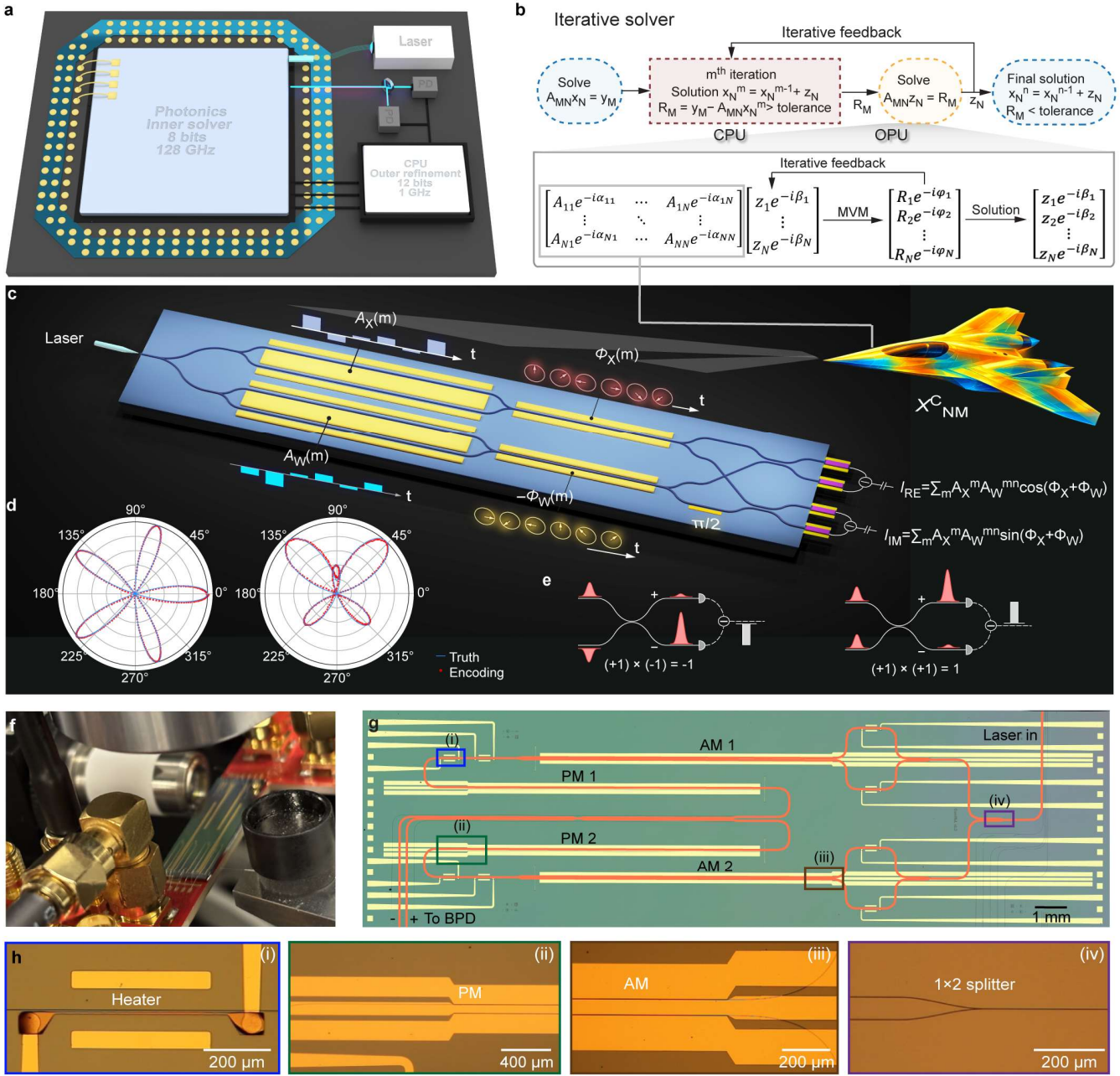

**Fig. 1|** Concept of mixed-precision architecture and TFLN computing device. **a.** Photonic-digital hybrid computing system. **b** Mixed-precision algorithm with iterative solver, where CPU performs critical refinement, and OPU accelerates the dominant workload by performing high-throughput matrix multiplications. **c.** Photonic device for amplitude and phase modulation for complex number multiplication using homodyne detection. Current distribution around an aircraft (figure credit: GPT-5) is an example of PDE solvers for RCSs. **d.** Experimental validation of amplitude and phase encoding of complex-valued 'flower' and 'butterfly' functions (red dots) with comparison to the ground truth (blue lines). **e**. Principle of homodyne computing logic based on coherent detection. **f.** The experimental setup of a wirebonded TFLN chip. **g** Fabricated TFLN chip with parallel amplitude and phase modulators. **h** Zoomed-in views of **i** heater for modulator bias, **ii** phase modulator electrodes, **iii** amplitude modulator electrodes, and **iv** input Y splitter.

Our architecture consists of a homodyne photonic device that performs high-speed complex-valued multiplications $Y_n^C = W_{mn}^C X_m^C$, where $X_m^C = A_X^m e^{-j\phi_X^m} = A_X^m \cos(\phi_X^m) + j A_X^m \sin(\phi_X^m)$ and $W_{mn}^C = A_W^{mn} e^{-j\phi_W^{mn}} = A_W^{mn}\cos(\phi_W^{mn}) + j A_W^{mn}\sin(\phi_W^{mn})$ are in the polar form, which is widely used in electromagnetic simulation (Fig. 1c-1d). In the fabricated device (Fig. 1f-1h), a laser (with frequency $f$) is split into two beams (labeled $X$- and $W$-beams): each propagates through an amplitude modulator and a phase modulator that impacts, respectively, synchronized amplitude $A_X^m$ and phase values $\phi_X^m$ on the $X$-beam, and amplitude $A_W^{mn}$ and the phase $-\phi_W^{mn}$ on the $W$-beam using $m$ time steps, resulting in the electric fields $E_X(m) = A_X^m e^{-j2\pi ft + j\phi_X^m}$ and $E_W(m) = A_W^{mn} e^{-j2\pi ft - j\phi_W^{mn}}$. The total of $n$ vectors is encoded sequentially for matrix-vector multiplication. The two beams are combined with a 50:50 coupler and recorded on the balanced photodetectors (BPDs) that calculate the real part $A_X^m A_W^{mn} \cos(\phi_X^m + \phi_W^{mn})$, and the imaginary part $j A_X^m A_W^{mn} \sin(\phi_X^m + \phi_W^{mn})$ by means of a $\pi/2$ phase difference (via thermal tuning), which can be simultaneously measured using a 90-degree optical hybrid [46]. In the Cartesian form, the generated photocurrents corresponding to the element-wise photoelectric multiplication are accumulated over $m$-steps using two separated low-speed BPDs or charge integrating receivers (Methods), yielding the real and imaginary parts:

$$I_n^{\mathbb{C}} = \sum_m A_X^m A_W^{mn} \cos(\phi_X^m + \phi_W^{mn}) + j \sum_m A_X^m A_W^{mn} \sin(\phi_X^m + \phi_W^{mn})$$

The time integration is performed with low-speed BPDs with switchable bandwidth 0.1~150 MHz, depending on the data rates and the vector lengths. Compared to the switch integrators with 10 μs discharging time [8,10,38], the low-speed photodetector offers a scalable solution with low discharging times (~6 ns), significantly reducing the latency. This architecture holds several key advantages for optoelectronic mixed-precision computing. It allows (1) scalable data mapping using temporal pulses, e.g., over 100 billion parameters per second (100 GS/s); (2) interferometric phase stability with all the key components (beam splitters, combiners, optical paths) on-chip, and (3) parallel amplitude and phase encoding without coupling crosstalk using TFLN EO modulation at CMOS-compatible voltages without amplification; (4) linear operations of signed number with amplitude response but not intensity response, and time integration readout of (5) high SNRs and low optical power, and (6) energy-efficient analog-to-digital conversion (ADC) circuits with high resolution; (7) reducing the optical cycles (>100 GS/s) to the electronic clock rates (0.1~1 GHz) via time integrated readouts; and (8) high throughput as each complex operation effectively equals 6 real-number operations.

## 2.2 Homodyne computing and accuracy optimization

We first verify the accuracy of amplitude and phase encoding. The TFLN modulators are designed with matching microwave and optical propagation indexes in a traveling-wave configuration, and the electrodes are terminated with a 50 Ω resistor for impedance matching. The modulators are aligned and oriented toward the edge for the ease of wirebonding. Another version of the chip with only amplitude modulators is fabricated for high-speed real-valued multiplications using electrical probes. The modulators exhibit a broadband electro-optic bandwidth over 40 GHz with a half-wave voltage of 2 V for 1.5 cm electrode length, similar to our previous work [7]. We verify the amplitude-phase precision by encoding a target complex pattern (flower or butterfly) on one of the two paths, and the homodyne readout reveals the real and the imaginary parts of the complex numbers on BPDs that recover the patterns at 10 MS/s. The result in Fig. 1d is compared with the digital ground truth, and the discrepancy exhibits statistical errors of σ=0.39% for the real part and σ=0.43% for the imaginary part, corresponding to a mean square error (MSE) of less than MSE=0.00002 and representing a precision of $B$=9 bits calculated by $B = \log_2(1/\sigma) + 1$, accounting 1 bit for the sign, representing excellent phase and amplitude precision. However, the statistical errors increase with higher clock rates (Table 1), as we observed σ=5.8% (MSE=0.0036) at 40 GS/s and σ=10% (MSE=0.01) for 128 GS/s, similar to the results in [6].

**Channel response equalization.** A key challenge to improve computing accuracy at high clock rates ($R$) is the requirement for uniform channel response across the bandwidth from DC to $R/2$, where subtle distortions, such as reflections (due to impedance mismatch), high-frequency attenuations, or transient oscillations, degrade the signal integrity. This requirement is demanding for high computing bits because the error tolerance scales exponentially $\sigma<1/2^{B-1}$. Here we develop frequency-domain equalization, where the channel response $H(f)$ is measured with homodyne detection as illustrated in Fig. 2a (Methods). Any unseen input signals are thus pre-compensated with the $1/H(f)$ filter before encoding. Note that the filter can be implemented with pre-emphasis electrical drivers for real-time correction [47]. Fig. 2b and 2c exemplified the temporal traces from the Modified National Institute of Standards and Technology (MNIST) database weight data with 28x28 pixels at 40 GS/s with comparison to its digital ground truth. The frequency response is calibrated up to 20 GHz, due to our BPD bandwidth cut-off limit. And the frequency domain response exhibits the effectiveness of our channel equalization, where the errors, by comparing the output signal over 784 symbols (Fig. 2d and 2e), are reduced from 6.0% (MSE=0.0036) to 1.6% (MSE=0.00025) with calibration, which corresponds to 7-bit quantization accuracy. The encoded data were transformed into the frequency domain to demonstrate the benefit of the frequency-response calibration as illustrated in Fig. 2f. The accuracy improvement is limited by the accuracy of the measured transfer function $H(f)$, for which we benefit from the high bandwidth and the linearity of TFLN modulators.

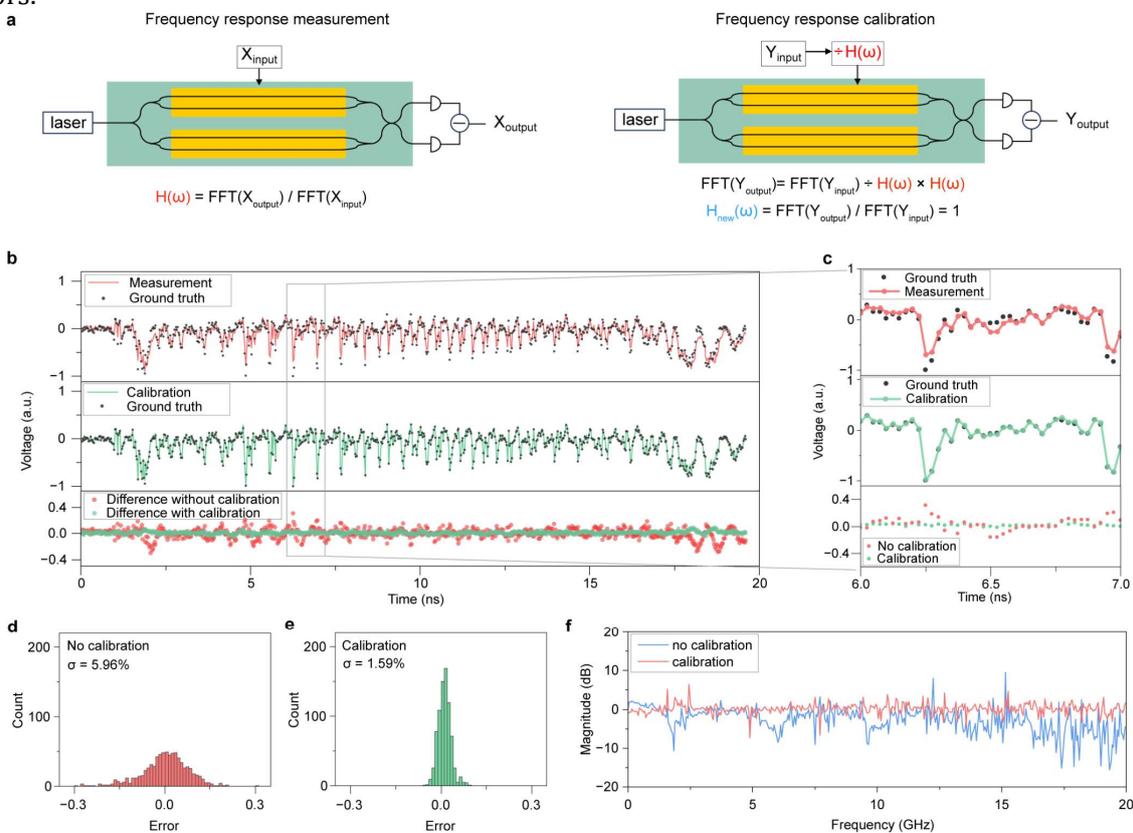

**Fig. 2|** frequency response calibration. **a** Left: channel response measurement by encoding random numbers and Fourier transform to get input-output response $H(f)$. Right: pre-compensation of any unseen input data by multiplying $H(f)$ in the frequency domain. **b-c** Temporal traces of MNIST weight encoding before (red) and after (green) calibration at a 40-GHz clock rate, showing improved accuracy. **d-e** Histograms of encoding errors, where calibration reduces the standard deviation from 5.96% (~5-bit precision) to 1.59% (~7-bit precision). **f** Corresponding frequency responses of the weight encoding, exhibiting large fluctuations without calibration (blue) and a much flatter spectrum after calibration (red).

**Benchmarking AI computing.**
We explore a two-layer complex number neural network in AI inference to recognize MNIST images (Fig. 3a and 3b). Parallel amplitude and phase modulation enable complex-valued weighting and readouts. The encoding clock rate was set to 100 MS/s due to the number of high-speed drivers available. We perform frequency response calibration to improve the encoding accuracy. The measured real and imaginary components are combined to form a complex output, and the magnitude of this complex value is used for final classification (Fig.

3c and 3d). The channel equalization reduces the discrepancies between the ground truth and the measurement errors from 2.97% to 1.69% (Fig. 3e-3h), which improves the classification accuracy from 90.1% (before calibration) to 93.4% (Fig. 3i-3k). This comparison highlights the effectiveness of the equalization in computing tasks.

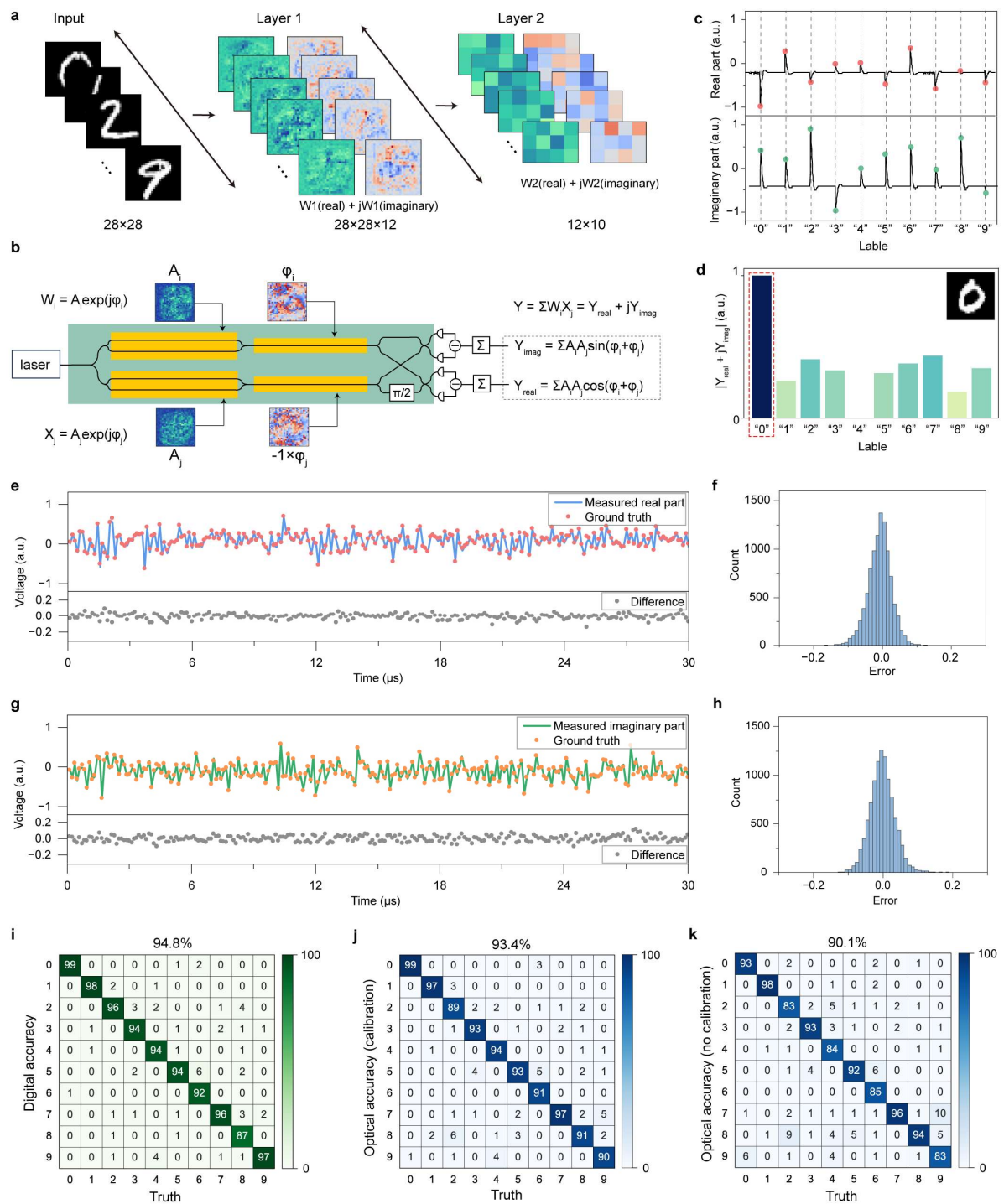

**Fig. 3|** Experimental results of complex-valued homodyne computing for MNIST classification. **a** Complex-valued ONN model with two layers. The 28×28 input images are mapped to layer 1 with 784×12 complex-valued weights (real and

imaginary parts), followed by layer 2 to generate 10 output features. **b** Principle of complex-valued VVM implemented on the TFLN photonic chip, where the multiplication of input $X_i$ and $W_i$ produces both real and imaginary parts through homodyne detection. **c** Example of measured integration temporal traces using a low-speed photodiode, showing real and imaginary channels. The baseline deviates from zero because of an offset in the photodetector readout. **d** Classification result of one representative input image. The measured real and imaginary outputs are combined in amplitude $(Re^2+Im^2)^{0.5}$ to form the output vector across 10 classes. **e** Measured real-part temporal trace of 250 VVM outputs (25 out of 1000 images) compared with ground truth, with the difference plotted below. **f** Error histogram of 10000 VVM outputs (1000 images) in the real part with a standard deviation of 1.69%. **g** Corresponding imaginary-part trace. **h** Errors histogram of the real part with a standard deviation of 1.76%. **i–k** Confusion matrices showing digital inference accuracy of 94.8%, optical accuracy of 93.4% with calibration, and optical accuracy of 90.1% without calibration.

We further benchmark the performance of our device for high-speed AI inference using amplitude modulators for real number operations, where we loaded a pretrained single-layer neural network (28x28→10) with digital accuracy of 91% to recognize MNIST images (see Methods). At the computing clock rate of 128 GS/s, each input image with 28x28=784 pixels multiplies with weight vectors (Fig. 4a) within a duration of τ=6.125 ns (Fig. 4c). The measured integrated temporal trace (Fig. 4b) agrees with the ground truth within a statistical error of 4.0% and it corresponds to an accuracy of about 6 bits at 128 GS/s, which is about 2 bits higher than that in [20]. We attribute the residual errors to the insufficient calibration bandwidth, as our TFLN modulators are calibrated only up to 22 GHz, which can be improved with BPDs of higher bandwidth. The classification performance for 102 unseen random image samples reaches 92.16% accuracy (Fig. 4d-4e), which is 96.9% of the digital accuracy of the same test dataset (95.10%), within the statistical errors.

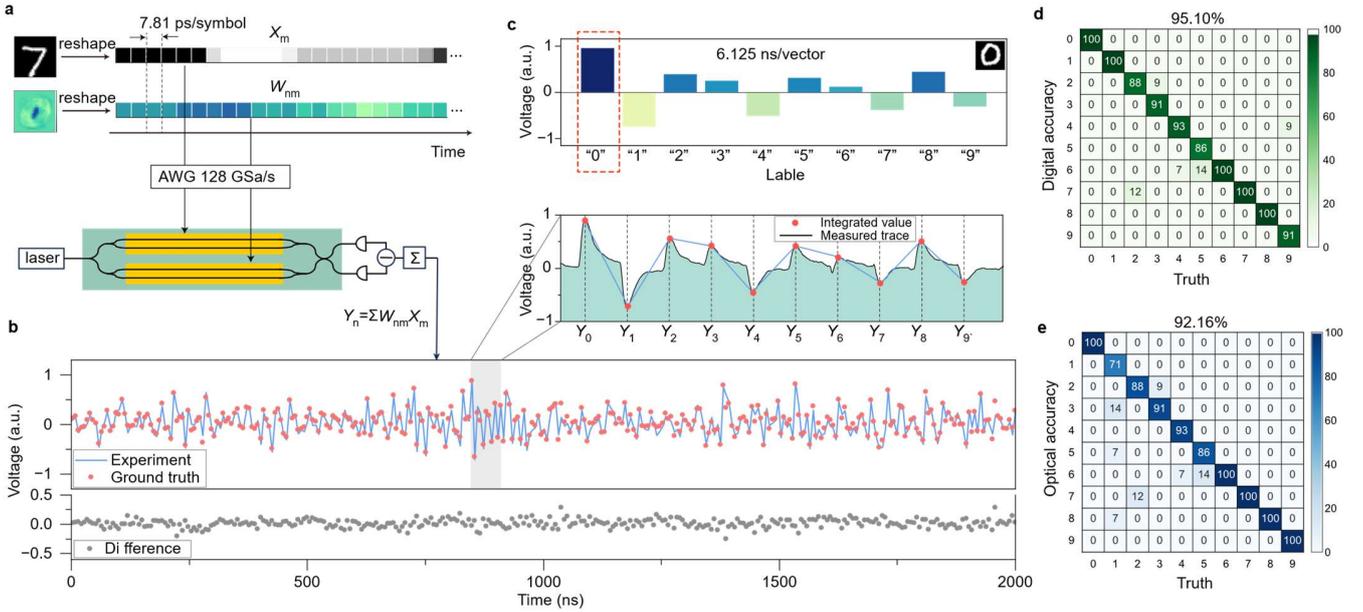

**Fig. 4|** Experimental classification of handwriting digits at 128 GS/s clock rate. **a.** Schematic of the high-speed VVM (vector-vector multiplication) for a single-layer ONN model. **b** Measured temporal trace of 326 VVM outputs (out of 1020 VVM outputs) with an inset corresponding to **c** the classification of one representative input image. **d–e** Confusion matrix comparison of the digital and optical accuracy.

## 2.3 Photonic-accelerated mixed-bit scientific simulation
### Quantization-aware iterative PDE solver

We implement quantization-aware algorithms to further improve optical computing accuracy, which is benchmarked in solving the charge density distribution on a thin, conducting wire [48,49]. This is a prototype problem for singular perturbations in electromagnetism, where it finds the boundary state required for a constant-potential harmonic function to satisfy on a finite, vanishingly thin domain, and is widely used in thin-wire antennas, near-field coupling in nanoscale conductors, quasi-static limits of plasmonic, and photonic structures. Our solver aims to compute the axial charge density distribution ρ(x) with 12-bit precision. The wire with a length $L$ and a radius $a$ is subjected to an applied static voltage $V_s$ (Fig. 5a). In the quasi-static limit, the electrostatic potential $V(\mathbf{r})$ satisfies Laplace's equation $\nabla^2 V(r, z, \theta) = 0$. The induced charge density $\rho(x)$ arises implicitly from enforcing the equipotential on the conductor surface, which using the Green's function representation, reduces the partial differential equation (PDE) to a boundary integral equation (BIE), $\frac{1}{4\pi\varepsilon_0} \int_0^L \frac{1}{\sqrt{(x-x')^2+a^2}} \rho(x')dx' = V_S$. This can be calculated with real-valued matrix-vector multiplication (MVM) $V(m) = \sum_{n=1}^{N} A(m,n)\rho(n)$, by discretizing the length $L$ is to $N$ segments with $A(m,n) = \frac{1}{4\pi\varepsilon_0} \int_0^{L/N} \frac{dx'}{\sqrt{(x'+(n-m)\frac{L}{N})^2+a^2}}$ describes the coupling strength between sites $m$ and $n$. Due to the high dynamic ranges of the coupling matrix, the digital model deviates from the ground truth even with 16-bit digital precision.

We thus implement an outer-inner loop preconditioned conjugate gradient (PCG) iterative solver (Fig. 5b) that calculates $\rho(n)$ with $\rho_N^{k+1}=\rho_N^k+z_N^k$ in two separate parts. In each $k$-th iteration, the digital outer loop verifies $V_M^k=A_{MN}\rho_N^k$ with the static potential $V_S$ by feeding the discrepancy $t_M^k=V_S-V_M^k$ to the low-precision photonic inner loop which calculates a correction term $z_N^k$ to satisfy $t_M^k = A_{MN}z_N^k$ (M=N=100 denoting the matrix size). Mathematically, the outer loop extracts the significant digits from the ground truth, which allows the discrepancies to be calculated at low accuracy in photonics. This hybrid digital–optical solver converges to a charge distribution within a set error tolerance of 0.1%. Experimentally, we encoded the data with amplitude modulation at the clock rate of 128 GS/s, given that the charge model is real-valued. The convergence rate of the solver (toward low residuals) is governed by the computing accuracy of the optical processor (Fig. 5c). Our solver converges with 2 outer iterations (Fig. 5h–5j), with the inner loop optical matrix outputs (Fig. 5d, 5e) and the correction vector $z_N^k$ (Fig. 5f, 5g). Comparing experiment convergence and digital simulations at different quantization precisions, we verify that our experimental precision is about 6 bits. The final solution agrees well with the ground truth within a standard deviation of 0.25% in the discrepancy. The entire solver operates 3 MVMs in the outer loop and 11 MVMs (9 operations with $A_{MN}p_N$ and 2 operations with $A_{MN}z_N$) in the inner loop, where the optical process accelerates 78.6% of the operations.

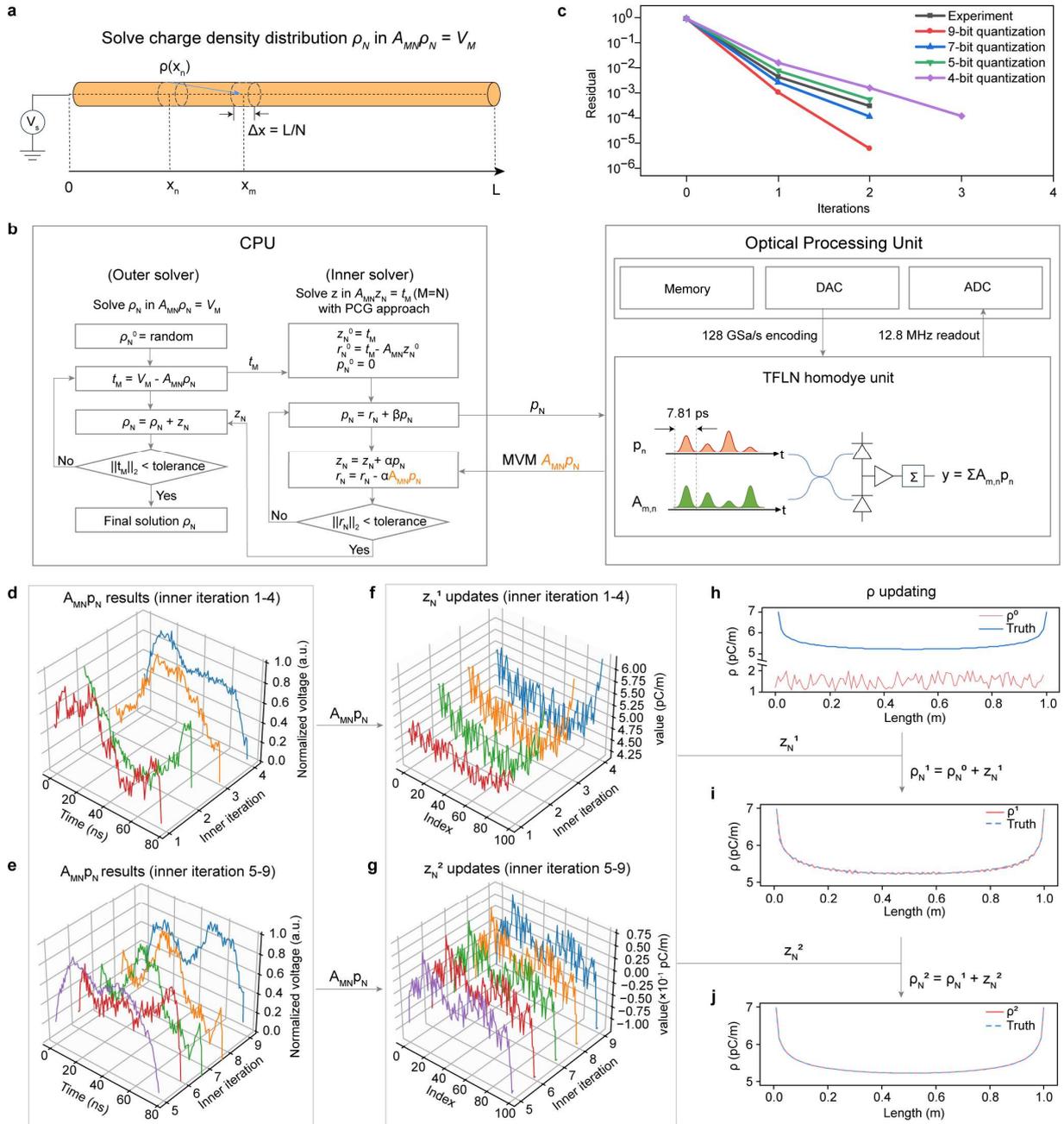

**Fig. 5|** Experimental demonstration of iterative PDE solver for charge distribution on a photonic processor at 128 GS/s clock rate. **a** Schematic diagram of the conductive wire. **b** Mixed-precision iterative refinement algorithm combining electronic and optical MVM. $A_{MN}$, coupling matrix; $\rho_N$, charge distribution; $V_M$, potential distribution; $t_M$, potential residual; $z_N$, charge correction (inner solver solution); $r_N$, residual of inner solver; $p_N$, conjugate search direction; $\alpha$, step size along the conjugate direction; $\beta$, conjugacy coefficient. **c** Experiment and simulation residual convergence rates versus iteration number. **d, e** Photonic MVM $A_{MN}p_N$ outputs of the inner solver to update the correction vector $z_N$ in each inner iteration. **f, g** Evolution of the correction vector $z_N$ computed by the inner solver. The computed MVM results $A_{MN}p_N$ iteratively update the correction vector $z_N$ till the residuals are below a specified tolerance. **h-j** The evolution of solution $\rho_N$ in the digital outer solver shows progressive agreement with the ground-truth distribution.

**Dense and sparse quantization in PDE solver**

Increasing the model complexity, we solve the complex-valued electromagnetic scattering (as current distribution) along a conducting wire under the excitation of an oscillating electric field e$^{j\omega t}$ (Fig. 6a). The system is modeled with the PDEs of integral form using free-space Green's function with an effective-distance treatment to regularize self and near-self interactions, as $jk_0 Z_0 \int_{-L}^{L}[I(z')G(z,z') + \frac{1}{k_0^2}I(z')\frac{\partial^2}{\partial z^2}G(z,z')]dz' = E_i(z)$, where $I(z')$ is the unknown electric current distribution, $G(z,z')$ is the reduced Green 3D function, $k_0$ is the wavenumber in free space, $Z_0$ is the free space wave impedance, and $E_i(z)$ is the incident field that can be modeled by a delta gap excitation. Discretizing the electric-field integral equation with the method of moments on a uniform 1D grid yields a dense complex system $A_{MN}I_N = b_M$, where $A_{MN}$ is an M×N coupling complex matrix (M = N = 101), $I_N$ is the current vector representing the induced current samples along the wire, and $b_M$ is a nonzero excitation vector. We thus solve the current distribution using the same inner-outer iterative solvers architecture as Fig. 5b using the Generalized minimal residual method [50]. However, the entries in the impedance matrix $A_{MN}$ vary dramatically in magnitude. While the near-diagonal elements are comparatively large due to strong local interactions, most off-diagonal elements are close to zero, creating a substantial dynamic-range imbalance. The uneven magnitude distribution poses a challenge for on-chip computation, in which the limited 8-bit resolution cannot faithfully represent both the dominant near-diagonal values and the much smaller off-diagonal ones simultaneously.

To further improve the dynamic range of the inner-loop iterative solver, we implement dense-sparse quantization, a mix-precision computing technique that is emerging in AI computing for weight compression [51,52]. We decompose the coupling matrix $A_{MN}$ into a sparse component $S_{MN}$ representing the large-magnitude near-diagonal terms, and a remainder $D_{MN}$ containing the low-magnitude off-band terms, $A_{MN} = S_{MN} + D_{MN}$ (Fig. 6d). The sparse part, which includes 21 diagonals accounting for approximately 20% of the total computational workload, is processed digitally. The dense complex-valued remainder matrix is loaded to the photonic chip with simultaneous amplitude and phase encoding (Fig. 6b), allowing the hardware to process the majority of operations (80%). Figure 6e shows an experimental example of $A_{MN}v_N$ evolution in the inner solver for the first 50 iterations. Starting with a zero initial solution, the model converges after 4 outer iterations corresponding to 200 mixed-precision MVM operations in the inner optical solver. Although the digital part $S_{MN}v_N$ dominates the overall magnitude, the dense contribution $D_{MN}v_N$ is critical for the convergence. The evolution and final solution are shown in Fig. 6c, exhibiting an excellent agreement with the digital ground truth and a statistical error of 0.2%.

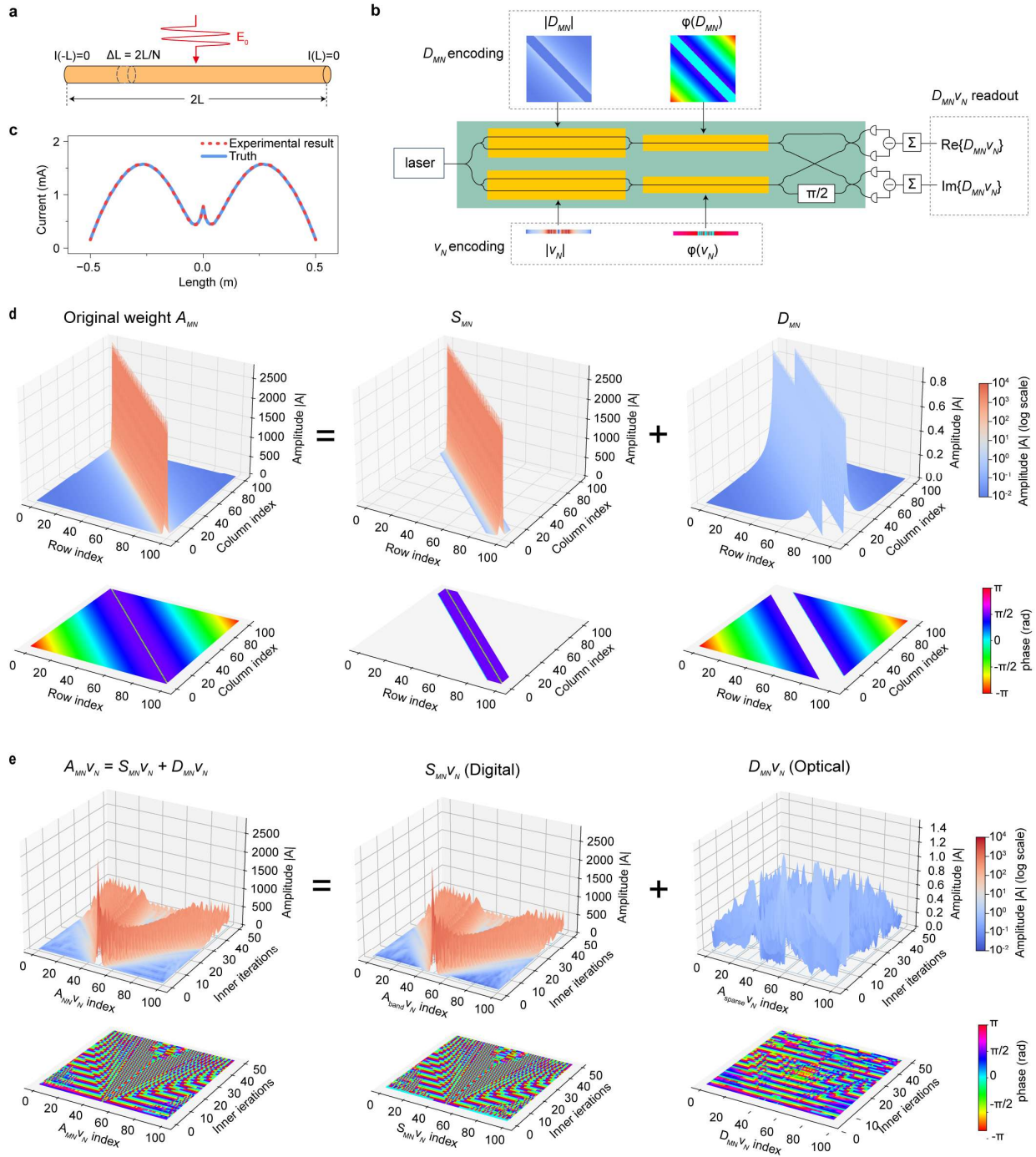

**Fig. 6|** Experimental demonstration of complex-valued iterative solver based on sparse-dense mixed-precision architecture. **a** Scattering problem with a thin-wire electric-field integral equation. A perfectly conducting wire of length $L$ is excited at its center by an incident field, and the induced current distribution is solved via the method of moments (MoM) on a uniform discretization. **b** Schematic of the complex-valued MVM for the inner solver. **c** Final solution of the current distribution. **d** The decomposition method of the complex-valued weight matrix $A_{MN}$ into a band component $S_{MN}$ and an off-band component $D_{MN}$. Top histogram is the amplitude distribution, and the bottom is the phase panel. **e** Mixed-precision MVM in GMRES inner solver. The digital processor computes the high-precision contribution $S_{MN}v_N$, and the photonic hardware evaluates the low-precision component $D_{MN}v_N$. The combination restores the full matrix–vector result $A_{MN}v_N$.

## Emulation of 3D models with bit-slicing quantization

To further generalize our mixed-precision framework toward complicated problems, such as three-dimensional (3D) EM simulation of radar cross-sections (RCS), we propose a bit slicing technology (Methods) to further improve the computing accuracy. The discretized electric surface current densities in the metallic aircraft have 15800 unknowns and the system matrix has 15800×15800=249.64 millions of elements (Methods). Using a digital processor, the solver converged after 122 iterations, requiring 122 slow MVMs performed in full precision (64-bit precision). Due to the complexity and the size of the model, this solver was implemented in a digital emulator that catches the main computing features of our homodyne computing device in the inner layer.

Fig. 7a and 7b illustrate the magnitude of the resulting equivalent electric current density. Therefore, applying the mixed-precision approach in our inner-outer iterative solvers in combination with the dense-sparse approach alone is not sufficient to compensate for the low accuracy of the optical MVM and the solver doesn't converge to a solution. Therefore, we quantize the system matrix and vectors to 16 bits, apply bit slicing, and perform each MVM as four 8-bit MVMs (Fig. 7c and Methods) that fit the computing accuracy of the homodyne processor. We manage to solve the problem using an 8-bit optical tensor model while requiring only 3 high-precision MVMs in the outer loop of the mixed-precision approach. We emulate the solutions when deploying them to the optical chip (Methods). The final resulting relative error of the total RCS is 0.0004, indicating that the algorithm has reached an accurate result. Fig. 7d and 7e show the total RCS error over the full $4\pi$ steradians, as well as the normalized RCS of the glider in the xz-plane, yz-plane, and xy-plane, for both the digital full-precision solver and the mixed-precision optical solver in each outer iteration, respectively. Although this approach increases the computational cost (Methods) of each iteration compared to a single 8-bit MVP, it remains significantly faster than using a 64-bit digital processor.

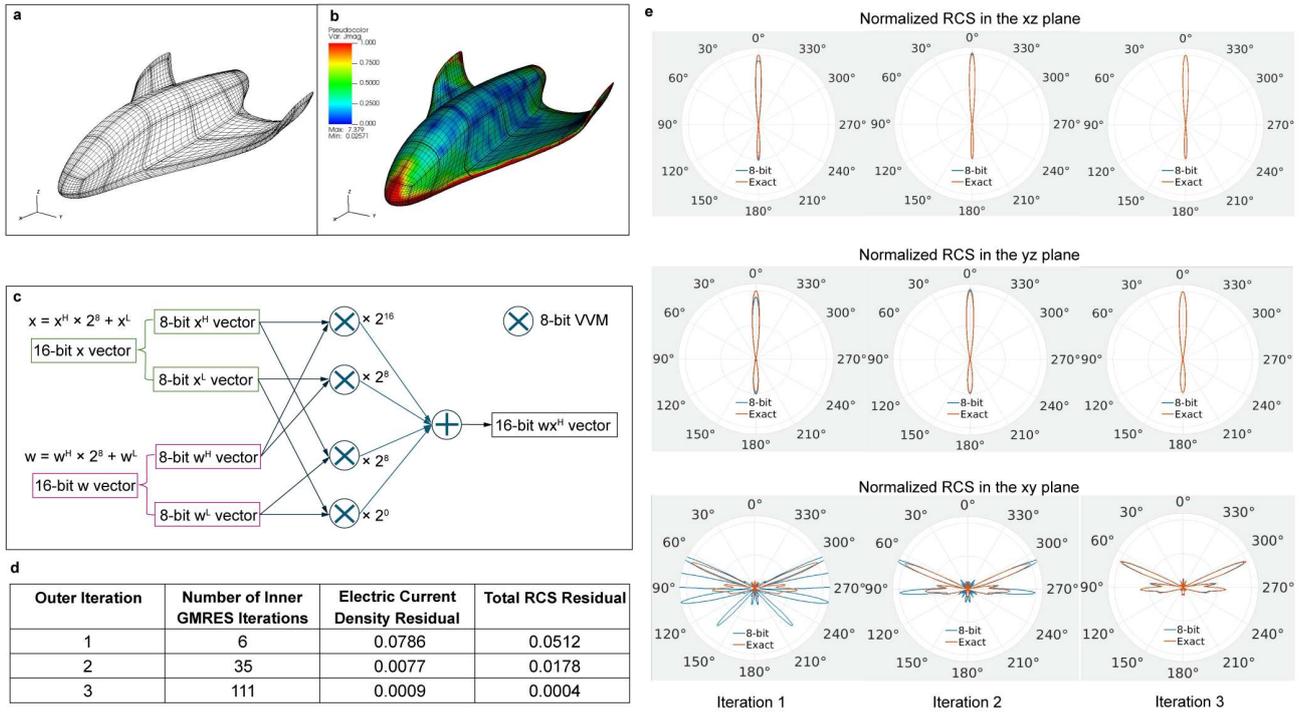

**Fig. 7| Glider scattering. a** Glider structure composed of surface patches. **b** Magnitude of the electric current density on the glider's surface. **c** Bit-slicing technique demonstration: two vectors *x* and *w* are decomposed into 8 bits, and then multiplied using 8-bit optical processors. **d** The details of the mixed-precision iterations, including the electric current density residual, and total 3D RCS residual. **e** The normalized RCS in xz, yz, and xy planes in each outer iteration.

**Discussion**

We demonstrated a quantization-aware optoelectronic framework that allows high-speed, low-accuracy analog photonic processors to accelerate high-precision computational tasks, which represents a significant step toward general-purpose computation with optics, for both AI computing and beyond. This enables, for the first time, optical computing with sufficient accuracy to directly solve scientific EM problems that would normally only converge with over 16-bit accuracy, using co-design architectures, including the inner-outer loop iterative solvers, the sparse-dense matrix decomposition, and the bit-slicing technology, to tackle problems of different complexity, resulting in solutions of a mean square error within a few parts per million MSE=0.000004 ($\sigma$=0.002). Secondly, we develop a linear homodyne computing logic, a key component to scale up photonic computing with its technical performance summarized below:

**Compute Accuracy.** Based on our channel equalization techniques, the computing accuracy reaches 6 bits at 128 GS/s in the EM real-number simulation, which is 2 bits better compared to the direct modulation results at the same speed [6]. The achieved accuracy can be improved further with a higher calibration bandwidth.

**Projected throughput.** We demonstrated a combination of real-number operations at high speeds and complex number operation at low speeds due to the limitation in high-speed drivers. In the near term, computing complex numbers at high speed may reach an effective throughput of $T$=2x6x128 GOPS=1.536 TOPS, considering 1 complex multiplication equals 6 real-number operations and a factor of two accounts for simultaneous multiplication and accumulation. With parallel processing with spatial fanout in a crossbar architecture [42,53] with over 100x100 channels, the throughput may be improved by a factor of 10,000 to the 15 POPS level, which would outperform the state-of-the-art GPUs by 10 times in a single chip.

**Latency.** Though the latency of each multiplication is limited by the time for encoding a vector, the latency, benefiting from the high-speed encoding, is as short as several nanoseconds. This latency is sufficiently low for many application scenarios such as autonomous driving or real-time analysis as RCSs.

**Projected energy efficiency**. The required energy cost, consisting of optical power and electronics to drive the devices, is estimated to be 520 mW (Methods), corresponding to 3 TOP/W, which is similar to that of a state-of-the-art GPU. As the analog-to-digital converters (ADCs) in our circuits are low power due to low-speed integrated readouts, the power consumption is dominated by the electronic driver, such as digital-to-analog converters at over 100 GHz for optical transceivers that achieved less than 1 pJ/bit conversion [54]. In the near term, the energy efficiency can be improved by 100-fold to the order of 10,000 TOP/W when incorporating parallel processing in a crossbar with over 100x100 channels [42,53] that rescues each converted data 100 times to about 250 TOPS/W. As TFLN modulators are driven with a voltage of only ±0.4 V, the energy consumption may be further reduced. Moreover, recent progress on optoelectronic digital-to-analog converters (ODACs) using TFLN modulators may further [55] reduce the electronics overhead to reach the modulator energy consumption at only 1.6 mW per device, and that corresponds to only ~1 fJ/data encoding at 128 GS/s for a potential computing efficiency of 200 TOPS/W.

In conclusion, quantization-aware photonic homodyne computing establishes a new paradigm in which ultrafast, low-precision optical hardware is no longer constrained to low-precision workloads, but can instead be systematically elevated to support high-fidelity, scalable digital simulations and artificial-intelligence computation through algorithm–hardware co-design. Looking forward, advances in high-speed electronic–photonic interfaces, broadband calibration, and co-packaged optoelectronic convertors [55] will further expand the attainable precision–throughput–energy trade space, enabling dense arrays of homodyne compute units to operate as reconfigurable accelerators with on-chip crossbars [42,53] for high throughput scaling. When combined with emerging mixed-precision algorithms, sparse and structured operators, and domain-specific solvers, such architectures offer a scalable path toward solving large-scale, complex-valued problems that are prohibitively expensive for purely electronic systems, such as real-time PDE solvers for radar cross sections and object detection. As quantization-aware mixed-precision algorithms are emerging in AI computing in large language models, where 99.5% computing tasks can be low precision [51], optical computing can offer a fast and efficient solution for these tasks. Beyond accelerating today's AI inference, by solving the fundamental partial differential equations, this approach opens opportunities for real-time inverse design, multiphysics modeling, and adaptive learning in wave-based systems, pointing toward a future where photonic processors become integral components of general-purpose, energy-efficient computing platforms, in which physical models and data-driven methods are tightly coupled, enabling scalable digital twins, uncertainty-aware sensing, and edge-deployable physics-based intelligence across applications ranging from remote sensing and autonomous systems to scientific computing and materials discovery.

## Methods

**Frequency response measurement**
To equalize the frequency response of each modulator, we first measure the transfer function of the channel that consists of an AWG, RF cables, a modulator, a homodyne detector and an oscilloscope. We send in sets of randomly distributed values at the target computing data rate from the AWG and record the output from the oscilloscope. The time domain data is Fourier transformed and divided by the spectrum of the ground truth. We extract the averaged transfer function from 100 datasets with 1000 random symbols. This transfer function is then applied in the pre-compensation part which adds a $1/H(\omega)$ filter to the unseen data before encoding. Since the highest-bandwidth balanced PD available in our system was 22 GHz (Optilab BPR-22-M), the maximum clock rate at which frequency-response calibration could be reliably performed was limited to 44 GS/s. The encoding calibration measurement at 40 GS/s employed Keysight M8199A AWG (maximum 130 GS/s).

**Time-integration.** We integrated the currents from the PDs using off-the-shelf low-speed time integrators (TI IVC 102 [8,10,38]), but that requires over 10 μs discharging time. The currents integrated with a BPD of matching analog bandwidth provide readout integration with fast discharging times for consecutive VVMs.

**Complex-valued VVM measurement**
Complex-valued VVM experiments were performed using a four-channel AWG (Tektronix AWG5014C). The half-wave voltage of amplitude and phase modulators is 2.6 V and 5.2 V. The amplitude modulators were biased at the intensity extinction point and driven with a 1.2 $V_{pp}$ voltage to ensure linear encoding. Any phase term could be mapped onto a $\pm\pi/2$ range, enabling arbitrary complex-valued encoding. Before applying the phase modulators, an integrated heater is used to bias the initial relative phase of the two arms. This configuration allowed simultaneous and accurate encoding of both amplitude and phase, thereby realizing full complex-valued readout of the integrated real part and imaginary part.

**Real-valued MVM measurement at 128 GS/s**
For the high-speed MNIST classification experiments, an AWG provides a 128 GS/s clock rate to drive amplitude modulations. The modulators were contacted using high-speed microwave probes to preserve bandwidth performance, while the integrated heaters were wirebonded to the PCB. This configuration ensured that bandwidth limitations from packaging did not constrain the 128 GS/s experiments. The readouts were performed with a 10 MHz photodetector.

**Aircraft radar cross-section.** The aircraft structure is illuminated by a plane wave propagating in z direction, and the magnetic field integral equation (MFIE) is solved to find the equivalent electric surface current densities on the surface. A Chebyshev-based boundary integral equation (CBIE) solver [56] is employed to decompose the object into surface patches, each discretized with *10 × 10* points, and find the equivalent electric surface current densities on each point, resulting in 15800 unknowns. The discretized system can be represented by a matrix equation, in which the system matrix has *15800 × 15800* elements. Generalized minimal residual method (GMRES), which is an iterative solver that performs an MVM in each iteration, is used to solve the system. As a result, this scattering problem is complex, with a large number of unknowns, in which a speedup in the MVM can significantly accelerate the solution time.
In our emulator, first, the dominant elements with magnitudes higher than the tolerance are multiplied separately (sparse-dense approach) with 16-bit precision, which reduces the dynamic range of the remaining matrix elements during the quantization process, resulting in lower quantization error. We set the tolerance so that only 0.003% of the data is separated; thus, their separate high-precision multiplication during each iteration introduces negligible computational overhead. Second, due to the complexity of the system being represented by only 8-bit values, we operate all the 16-bit MVMs with bit slicing that breaks it down into four 8-bit MVM operations. The inner GMRES tolerance is set to 0.1 to achieve convergence.

**Inference of AI models.**
In the complex-valued model, the input layer from an image consisting of 28x28 pixels flatten to 784 time steps feeding to layer 1 with a pretrained matrix of 784x12 parameters (Fig. 3a). The computing results from layer 1 are serialized and fed to computing layer 2 as the chip inputs (Fig. 3b). The temporal traces of both the real and imaginary outputs are measured with the balanced photodiode. These complex-valued outputs are then converted into the intensity and phase form, and only the intensity is used as the final output for classification across the 10 classes.

**Bit slicing.** Each 16-bit vector $X_N$ and $Y_N$ is separated to two 8-bit data by $X_N = X^H \times 2^8 + X^L$ and $W_N = W^H \times 2^8 + W^L$. So the multiplication of two bit-sliced vectors is computed with the weighted sum of 4 effective 8-bit multiplications, $Y_N = X_N W_N = \Sigma_N(X^H W^H \times 2^{16} + (X^H W^L + X^L W^H) \times 2^8 + X^L W^L)$. When deploying to the analog optical computing, to avoid analog noises in the $\Sigma_N X^H W^H$ term dominating the other terms, although it's digitized with a 8 bit, the noise floor should be maintained low for a 16-bit signal-to-noise ratio [57], which is achievable in our architecture due to the low-speed integration readout. For intermediate computing bits, for instance, a 12-bit data can be sliced to optoelectronic with 4-bit digital $\Sigma_N X^H W^H$ and 8-bit analog for the lower terms.

**Projected power consumption**
The energy consumption includes both optical and electrical power with two amplitude modulators and two phase modulators at the clock rate of 128 GS/s. The laser source consumed 1 mW. For data encoding, the energy consumption of the TFLN modulators can be estimated from their driving conditions. The amplitude modulators were driven in the quasi-linear region with a peak-to-peak voltage of 0.83 V, corresponding to a root-mean-square drive voltage of 0.29 V. With a 50-Ω termination, this results in an energy cost of approximately 13 fJ per operation using two amplitude modulators. In a full system, we consider each modulator is driven with a digital to analog converter with 1 pJ/conv, which dominates the total power at 519.9 mW. This corresponds to an energy efficiency of about 3 TOPS/W. Further development on optical digital to analog converter might reduce the driver power to the fundamental TFLN modulator power consumption, where the total power of the system is reduced to only 8 mW, corresponding to about 200 TOPS/W.

**Data availability**
The data that support the findings of this study are available from the corresponding author upon reasonable request.

**Acknowledgements**
This work was supported by the DARPA NaPSAC program under project N66001-24-2-4002. Z.C. and M.Y. acknowledge the Optica challenge awards. program, the 2025 Coherent II-VI foundation award, The National Science Foundation ASCENT program, and the Berkeley Baker fellow program.

**Author contributions.**
Z.C. conceived the project. L.Z., K.X., and L.J. conducted the experiment, assisted by Y.L., Y.L. and Y.Y.. C.H. fabricated the TFLN device, assisted by K.X. K.K. and C.C.. A.F. and C.S. developed the EM simulation algorithms with mixed-precision computing. S.Z., K.X., and L.Z. created the software model for neural network training. R.H., E.W., and D.E. provided experimental support and discussions. Z.C., M.Y and C.S supervised the project. Z.C. wrote the manuscript with contributions from all authors.

**Competing interests.**
Z.C, R.H, and M.Y are cofounders of Opticore Inc. and hold equity. D.E. serves as scientific advisor to and holds equity in Lightmatter Inc and Opticore Inc. Other authors declare that they have no other competing interests.

**Table 1. Random data multiplication at different speeds.**

| Datarates | Before calibration | After calibration | Instruments bandwidth |
|---|---|---|---|
| 50 MS/s | 0.78% (8 bit) | 0.25% (9~10 bit) | AWG: 9.8 GHz<br>OSC: 13 GHz<br>BPD: 22 GHz |
| 0.5 GS/s | 5.17% (5 bit) | 0.63% (~9 bit) | |
| 5 GS/s | 8.6% (4 bit) | 1.8% (7 bit) | |
| 40 GS/s | 5.68% (5 bits) | 1.72% (7 bit) | AWG: 63 GHz<br>OSC: 80 GHz<br>BPD: 22 GHz |

**Table 2. Comparison of computational precision and clock rate across photonic computing platforms.**

| Reference | Material platform | Compute logic | Clock rate | accuracy |
|---|---|---|---|---|
| Xu 2021 [5] | Bulk LN | Cascade intensity modulation | 62.9 GS/s | ~4 bits |
| Mourgias-Alexandris 2022 [31] | Silicon | | 10 GS/s | 3.07 bits |
| Lin 2024 [8] | TFLN | | 60 GS/s | 6 bits |
| Zhang 2025 [58] | TFLN | | 16 GS/s | 5.57 bit |
| Ou 2025 [7] | VCSEL + TFLN | | 10 GS/s | 5 bits |
| Al-Kayed 2025 [20] | TFLN | | 106 GS/s | 3.3 bit |
| This work with equalization | TFLN | Homodyne modulation | 10 GS/s | 8 bits |
| | TFLN | | 40 GS/s | 7 bits |
| | TFLN | | 128 GS/s | 6 bits |
| This work with mixed-bit computing | TFLN | | 128 GS/s | 12 bits |